# Whistler instability in a semi-relativistic bi-Maxwellian plasma

M. F. Bashir[1,2], S. Zaheer[3], Z. Iqbal[1,2] and G. Murtaza[1]

1: Salam Chair in Physics , G. C. University Lahore.
2: Department of Physics, G. C. University Lahore.
3: Department of Physics, F. C. C. University, Lahore.

**Abstract**

Employing linearized Vlasov-Maxwell system, the Weibel instability embedded in an ambient magnetic field is discussed for a semi-relativistic bi-Maxwellian distribution hoping such a scenario occurs in some relativistic environments e.g., gamma-ray burst sources and relativistic jet sources, supernovae, and galactic cosmic rays where the perpendicular temperature is much dominated over the parallel . The dispersion relations are analyzed analytically along with the graphical representation and the estimates of the growth rate are presented along with the instability threshold condition in the limiting cases i.e., $\xi_{\pm} = (\omega \pm \Omega)/k_{\parallel} v_{t_{\parallel}} \leq 1$ (resonant case) and $\xi_{\pm} >> 1$ (non-resonant case). It is observed that the relativistic effect suppresses the instability and also lowers the threshold for the instability to set in. The ambient magnetic field contribution to instability appears only in non-resonant case resulting in reduction of growth rate. However, the effect of the ambient magnetic field is diminished as we go from the weak relativistic regime to the highly relativistic one. We also note that the ambient magnetic field generates real oscillations and the reletivistic effect reduces these oscillations. Further for field free case i.e., $B_0 = 0$, the growth rates for Weibel instabilities are also presented in a semi-relativistic bi- Maxwellian plasma for both the limiting cases.

# I. Introduction

The presence of relativistic electrons in the magnetosphere induces various types of electromagnetic instabilities due to anisotropy of temperature. This instability arises in a variety of plasmas including fusion plasmas, both magnetic and inertial confinement, as well as in plasma created by highly intense free electron x-ray laser pulses. The classical Weibel Instability is such an example of an unmagnetized plasma and in the presence of magnetic field, the instability generated is either whistler or cyclotron maser. The Weibel instability, which has been around for several decades is of significant interest since it generates quasi-stationary magnetic fields, which can account for seed magnetic fields in laboratory and astrophysical plasmas.

The classical electromagnetic Weibel instability[1] in an unmagnetized plasma has the possible importance for generating cosmic magnetic fields in gamma-ray burst sources and relativistic jet sources, supernovae, and galactic cosmic rays[2-7] as well as the origin of cosmological seed magnetic fields in regions of intense gaseous streaming or temperature anisotropies[8-11]. By using different distributions, the anlysis of relativistic Weibel instability has been discussed in detail by several authors[12-20] A compartive study of Weibel and filamentation instabilities and their cumulative effects has been presented for non-relativistic and weakly relativistic bulk velocities by Lazar et al.[21] and Stockem et al.[22] respectively. Lately, the Weibel instability in quantum plasma has also been studied in linear regime by Haas[23] and in non-linear regime by Haas et al..[24]

The Weibel mode in a magnetized plasma i.e., the whistler wave is an electromagnetic wave in magnetized plasmas at frequencies below the cyclotron frequency $\omega_{pe} < \Omega_e$. Whistler mode emissions were detected inside and outside the Saturn's magnetosphere by plasma wave instruments on Voyager 2 [25, 26]. Whistlers are naturally produced in thunderstorms, lightning discharges and also near the north pole which can travel to the south pole along the Earth's magnetic lines of force through the Ionosphere and then return back to the origin. In magnetospheres whistlers are also observed to propagate through self created ducts.[27] In laboratory plasma, whistler mode is used for rf plasma discharge, heating of plasmas in tokamaks[28] and spheromaks[29]. Whistler instability in relativistic regime is a powerful mechanism for producing non-thermal, stimulated radiations (i.e., radio emissions)[30]. The necessary condition for this instability is that the positive gradient along perpendicular velocity should be present in velocity distribution function and such may occur in Solar corona[32], quasi perpendicular shocks[32] and the magnetosheath[33]. The most intense radiations originate from the strongly magnetized auroral regions of the magnetospheres, where the local electron plasma frequency is much less than cyclotron frequency. Such regions are also associated with other planetary magnetospheres and auroras [31-34].

The whistler instability was investigated for Waterbag distribution by Yoon and Davidson[13] and later by Achterberg and Wiersma[19] .Yang et al.[35] calculated the Weibel instability in a relativistic hot magnetized electron–positron plasma and showed that both the decrease in temperature anisotropy and increase in background magnetic field can cause a significant decrease in the growth rate. Davidson et al.[15] executed the stability analysis of the Weibel instability generated due to relativistic electron beam and compared it to Harris-type instability. Shah and Jain[36] studied the excitation of the whistler mode waves propagating obliquely to the constant and uniform magnetic field in a warm and inhomogeneous plasma in the presence of an inhomogeneous beam of suprathermal electrons. The full dispersion relation including electromagnetic effects is derived. In the electrostatic limit the expression for the growth rate is given. It is found that the inhomogeneities in both beam and plasma number densities affect the growth rates of the instabilities. Recently, Lazar et al.[49] discussed the Weibel instability in a magnetized non-relativistic bi-Maxwellian plasma and investigated the threshold conditions for the instabil-ity to set in. Mace and Sydora[37] investigated the parallel-propagating whistler instability in a magnetized plasma of electrons and positive ions having bi-kappa velocity distributions for a wide range of parameters. Liu et al.,[38] and Gary et al[39] presented linear kinetic dispersion analysis and performed a two-dimensional electromagnetic particle-in-cell simulation to demonstrate a possible excitation mechanism of whistler waves. Schlickeiser et al.[40] discussed the whistler Weibel-like modes in an anisotropic bi-Maxwellian magnetized electron-proton plasma.

Two of us ( i.e., Zaheer and Murtaza) recently discussed the Weibel instability for the non-Maxwellian distribution functions[41] and for the semi-relativistic Maxwellian distribution function[42] in an unmagnetized plasma. That work was later extended to a magnetized non-relativistic non-Maxwellian plasma.[43] More recently, two of us[44] (i.e., Bashir and Murtaza) presented a review study of plasma waves and insabilities and described the effect of temperature anisotropy on resonant and non-resonant whistler and Weibel instabilities for non-relativistic plasma. In the present paper, we investigate the whistler instability in the magnetized anisotropic plasma in the semi-relativistic Maxwellian regimes.

The layout of this paper is as follows: In section II, we use the kinetic theory to calculate the general dispersion relation for a magnetized plasma in both the non-relativistic and the semi-relativistic regimes using anisotropic Maxwellian distributions . We also derive the analytical expressions for the real and the imaginary parts of the dielectric constant for both the momentum distributions under the limiting cases $\xi_{\pm} \leq 1$ and $\xi_{\pm} \gg 1$. A brief summary of results and discussion is given in section III along with the graphical representation of the Weibel instability in a magnetized semi-relativistic bi-Maxwellian plasma for both the limiting cases.

## II. Mathematical Model

The linear dispersion relation for the transverse electromagnetic electron waves propagating parallel (i.e., $\mathbf{k}(0, 0, k_z)$) to the ambient magnetic field $B_0$, is given by[18]

$$\omega^2 - c^2 k^2 + \pi \omega_{pe}^2 \int_{-\infty}^{\infty} dp_{\|} \int_0^{\infty} dp_{\perp} \frac{p_{\perp}^2}{\gamma \left(\omega - \frac{k_{\|} p_{\|}}{\gamma m} \mp \Omega\right)} \left\{ \left(\omega - \frac{k_{\|} p_{\|}}{\gamma m}\right) \frac{\partial f_0}{\partial p_{\perp}} + k_{\|} v_{\perp} \frac{\partial f_0}{\partial p_{\|}} \right\} = 0 \quad (1)$$

where $\Omega = \Omega_0 / \gamma$ is the relativistic cyclotron frequency with $\Omega_0 = eB_0/m_o c$ and $\gamma^2 = 1 + p_{\perp}^2/m^2c^2 + p_{\|}^2/m^2c^2$ and $B_0$ is along the z-direction. In Eq. [1], the upper and lower signs in the denominator of the integrand correspond to the right hand and the left hand circular polarizations, respectively.

In the following, we shall derive the general linear dispersion relations for the non-relativistic and the semi-relativistic Maxwellian momentum distributions[42] i.e.,

$$f_0^N = \frac{1}{2\pi m T_{\perp}} \frac{1}{\sqrt{2\pi m T_{\|}}} \exp\left[-\frac{p_{\perp}^2}{2mT_{\perp}} - \frac{p_{\|}^2}{2mT_{\|}}\right] \quad (2)$$

$$f_0^S = \frac{\exp\left[\frac{mc^2}{T_{\perp}}\right]}{2\pi m T_{\perp} \sqrt{2\pi m T_{\|}} \left(1 + \frac{T_{\perp}}{mc^2}\right)} \exp\left[-\frac{mc^2}{T_{\perp}} \sqrt{1 + \frac{p_{\perp}^2}{m^2c^2}} - \frac{p_{\|}^2}{2mT_{\|}}\right] \quad (3)$$

Yoon[17] has studied the weibel instability with the fully relativistic anisotropic distribution function which in the limit of non-relativistic parallel momentum (i.e., $p_{\|}^2 \ll m^2c^2$ and $p_{\|}^2 \ll p_{\perp}^2$) gives semi-relativistic distribution function chosen above. We therefore assume that for both the non-relativistic and semi-realtivistic cases, the parallel momentum distributions are same having the non-relativistic Maxwellian distribution and the relativistic mass factor only depends upon the perpendicular momentum i.e., $\gamma \approx \gamma_{\perp} = \left(1 + p_{\perp}^2/m^2c^2\right)^{\frac{1}{2}}$. For $p_{\perp}^2 \ll m^2c^2$ and $T_{\perp} \ll mc^2$, the semi-relativistic distribution immideately reduces to the non-relativistic bi-Maxwellian distribution.

Thus performing straight forward $p_{\|}$-integrations with $f_{0\|} = \frac{1}{\sqrt{2\pi m T_{\|}}} exp[-\frac{p_{\|}^2}{2mT_{\|}}]$, Eq.(1) takes the form

$$0 = -\omega^2 + c^2 k_{\|}^2 - 2\pi \frac{\omega_{p_e}^2}{m^2 v_{t\perp}^2} \int_0^{\infty} \frac{dp_{\perp} p_{\perp}^3 f_{0\perp}}{\gamma_{\perp}^2} \left\{ \frac{\omega}{k_{\|} v_{t_{\|}}} Z(\xi_{\pm}) - \frac{1}{2} \left(\frac{T_{\perp}}{T_{\|}} - 1\right) Z'(\xi_{\pm}) \right\} \quad (4)$$

where

$$f_{0\perp}^N = \frac{1}{2\pi m T_{\perp}} \exp\left[-\frac{p_{\perp}^2}{2mT_{\perp}}\right]$$

$$f_{0\perp}^S = \frac{\exp\left[\frac{mc^2}{T_{\perp}}\right]}{2\pi m T_{\perp} \left(1 + \frac{T_{\perp}}{mc^2}\right)} \exp\left[-\frac{mc^2}{T_{\perp}} \sqrt{1 + \frac{p_{\perp}^2}{m^2c^2}}\right]$$

are the perpendicular distribution functions for the non-relativistic and the semi-relativistic cases respectively and $Z(\xi^{\pm})$ is tthe plasma dispersion function[46] defined as

$$Z\left(\xi_{\pm}\right) = \frac{1}{\sqrt{\pi}} \int_{-\infty}^{\infty} \frac{dS \ e^{-S^2}}{(S - \xi_{\pm})} \quad \text{with } \xi_{\pm} = \frac{|\omega \pm \Omega|}{k_{\parallel} v_{t_{\parallel}}} \text{ and } v_{t_{\parallel}} = \sqrt{\frac{2T_{\parallel}}{m}}$$

$Z'\left(\xi^{\pm}\right)$ indicates the derivative of plasma dispersion function with respect to its argument $\xi_{\pm}$.

By using the expansion of the plasma dispersion function for the limiting case $\xi_{\pm} \leq 1$

$$Z(\xi_{\pm}) \simeq i\sqrt{\pi} - 2\xi_{\pm} + \frac{4}{3}\xi_{\pm}^3 - ....$$

we may write the dispersion relation for R-wave in Eq.[2] as

$$0 = -\omega^2 + c^2 k_{\parallel}^2 - 2\pi \frac{\omega_{pe}^2}{m^2 v_{t\perp}^2} \int_0^{\infty} \frac{dp_{\perp} \ p_{\perp}^3 f_{o\perp}}{\gamma_{\perp}^2} \left[ \left(\frac{T_{\perp}}{T_{\parallel}} - 1\right) + i\sqrt{\pi} \frac{\omega}{k_{\parallel} v_{t_{\parallel}}} \left\{ \frac{T_{\perp}}{T_{\parallel}} - \frac{\Omega_0/\gamma_{\perp}}{\omega} \left(\frac{T_{\perp}}{T_{\parallel}} - 1\right) \right\} \right]$$
(5)

Similarly using the asymptotic expansion of $Z(\xi_{\pm})$ for the other limiting case $\xi_{\pm} \gg 1$. i.e.,

$$Z(\xi_{\pm}) \simeq -\frac{1}{\xi_{\pm}} \left(1 + \frac{1}{2\xi_{\pm}^2} + ....\right)$$

the dispersion relation for R-wave from Eq.[2] becomes

$$0 = -\omega^2 + c^2 k_{\parallel}^2 - 2\pi \frac{\omega_{pe}^2}{m^2 v_{t\perp}^2} \int_0^{\infty} \frac{dp_{\perp} \ p_{\perp}^3 f_{0\perp}}{\gamma_{\perp}^2} \left( \frac{\omega}{\omega - \Omega_0/\gamma_{\perp}} + \left(\frac{T_{\perp}}{T_{\parallel}} - 1\right) \frac{k_{\parallel}^2 v_{t_{\parallel}}^2}{2 \left(\omega - \Omega_0/\gamma_{\perp}\right)^2} \right)$$
(6)

We have used the expansion of the plasma dispersion function under the two limiting cases $\xi_{\pm} \leq 1$ and $\xi_{\pm} \gg 1$, the former case represents the resonant mechanism while the latter represents the non resonant one. For semi-relativistic case, the dispersion relations in Eq.(6) and Eq. (7) contain the relativistic effect through perpedicular Lorentz factor .

In the resonant case where the pole exists in the denominator, we require the collisionless absorption mechanism known as the cyclotron damping. On the other hand, for $\xi_{\pm} \gg 1$ i.e., the case of non-resonant phenomena we neglect the small imaginary contribution from the pole. Such wave particles interactions are very important in the heliosphere of Sun and magnetospheres of stars and galaxies. They may also occur in the environments where the effective wave frequency is large or small with respect to the thermal velocity of electrons i.e., $\left|(\text{Re}(\omega) \pm \Omega)/k_{\parallel}\right| \gg v_{t_{\parallel}}$ and $\left|(\text{Re}(\omega) \pm \Omega)/k_{\parallel}\right| \ll v_{t_{\parallel}}$ respectively.

In the following, we

## A.  Non-relativistic bi-Maxwellian Distribution

For the non-relativistic case, we use $f_{o\perp}$ from Eq.[3] and take $\gamma = 1$ in Eqs.[5] & [6].

Performing the perpendicular integration for $\xi_{\pm} \ll 1$, we may re-write the dispersion relation for R-wave from Eq.[5] as

$$0 = -\omega^2 + c^2 k_{\parallel}^2 - \omega_{pe}^2 \left[ \left(\frac{T_{\perp}}{T_{\parallel}} - 1\right) + i\sqrt{\pi} \frac{\omega}{k_{\parallel} v_{t_{\parallel}}} \left\{ \frac{T_{\perp}}{T_{\parallel}} - \frac{\Omega_0}{\omega} \left(\frac{T_{\perp}}{T_{\parallel}} - 1\right) \right\} \right]$$
(7)

For the subluminal case i.e., $|\omega| \ll ck$, the real and imaginary parts of $\omega$ are

$$\operatorname{Re}\omega = \frac{\Omega_0}{\frac{T_\perp}{T_\parallel}}\left(1 - \frac{T_\parallel}{T_\perp}\right) \tag{8}$$

and

$$\operatorname{Im}\omega = \frac{k_\parallel v_{t_\parallel}}{\sqrt{\pi}}\left(\frac{T_\parallel}{T_\perp}\right)\left[\left(\frac{T_\perp}{T_\parallel} - 1\right) - \frac{c^2 k_\parallel^2}{\omega_p^2}\right] \tag{9}$$

The instability occurs for waves satisfying the wave number condition

$$k_\parallel^2 < \frac{\omega_p^2}{c^2}\left(\frac{T_\perp}{T_\parallel} - 1\right) \tag{10}$$

Thus we observe that the magnetic field generates real oscillations but does not effect the growth rate. These non-relativistic results are the same as given earlier by Lazar et al[49] and Bashir and Murataza[44].

As is evident from Eqs. [8] & [9], in the absence of the ambient magnetic field, the non-relativistic mode becomes purely growing Weibel mode as expected. On ignoring the temperature anisotropy, we observe that the magnetic field effect vanishes and we obtain a purely damping mode.

For $\xi_\pm \gg 1$, we may re-write Eq.[6] after performing the perpendicular integration as

$$0 = -\omega^2 + c^2 k_\parallel^2 + \omega_{pe}^2\left(\frac{\omega}{\omega - \Omega_0} - \left(1 - \frac{T_\perp}{T_\parallel}\right)\frac{k_\parallel^2 v_{t_\parallel}^2}{2(\omega - \Omega_0)^2}\right)$$

Assuming $|\omega| \ll ck_\parallel$, we obtain

$$0 = \left(c^2 k_\parallel^2 + \omega_p^2\right)(\omega - \Omega_0)^2 + \Omega_0 \omega_p^2(\omega - \Omega_0) + \left(\frac{T_\perp}{T_\parallel} - 1\right)\frac{k_\parallel^2 v_{t_\parallel}^2 \omega_p^2}{2}$$

Solving the quadratic equation, we obtain the real and imaginary parts of $\omega$ as

$$\operatorname{Re}\omega = \Omega_0\left(1 - \frac{\omega_p^2}{2\left(c^2 k_\parallel^2 + \omega_p^2\right)}\right) \tag{11}$$

and

$$\operatorname{Im}\omega = \omega_p \frac{\sqrt{2\left(\frac{T_\perp}{T_\parallel} - 1\right)k_\parallel^2 v_{t_\parallel}^2\left(c^2 k_\parallel^2 + \omega_p^2\right) - \Omega_0^2 \omega_p^2}}{2\left(c^2 k_\parallel^2 + \omega_p^2\right)} \tag{12}$$

in agreement with the results of Lee,[47] Lazar et al.[49] and Bashir and Murtaza.[44]

Now the instability occurs for wave numbers satisfying the condition

$$k_\parallel^2 > k_m^2 = \frac{\omega_p^2}{2c^2}\left[\left\{1 + \frac{2}{\left(\frac{T_\perp}{T_\parallel} - 1\right)}\left(\frac{\Omega_0^2}{\omega_p^2}\right)\left(\frac{c^2}{v_{t_\parallel}^2}\right)\right\}^{\frac{1}{2}} - 1\right] \tag{13}$$

From Eq. [12], we observe that the growth rate is reduced due to the ambient magnetic field. In the limit of large temperature anisotropy i.e., $\frac{T_\perp}{T_\parallel} \gg 1$, the dispersion relation reduces to[47, 49]

$$\text{Im}\,\omega = \omega_p \frac{\sqrt{2k_\parallel^2 v_{t_\perp}^2 \left(c^2 k_\parallel^2 + \omega_p^2\right) - \Omega_0^2 \omega_p^2}}{2\left(c^2 k_\parallel^2 + \omega_p^2\right)} \tag{14}$$

where

$$v_{t_\perp} = \sqrt{\frac{2T_\perp}{m}}$$

On ignoring the magnetic field, the dispersion relation further simplifies to[1]

$$\text{Im}\,\omega = \frac{k_\parallel v_{t_\perp}}{\sqrt{2}} \frac{\omega_p}{\sqrt{\left(c^2 k_\parallel^2 + \omega_p^2\right)}} \tag{15}$$

From Eq. [12], we also note that for shorter wavelengths the growth rate becomes

$$\text{Im}\,\omega = \frac{\omega_p}{2} \sqrt{2\left(\frac{T_\perp}{T_\parallel} - 1\right) \frac{v_{t_\parallel}^2}{c^2} - \frac{\Omega_0^2 \omega_p^2}{\left(c^4 k_\parallel^4\right)}} \tag{16}$$

while for longer wavelengths

$$\text{Im}\,\omega = \frac{\sqrt{2\left(\frac{T_\perp}{T_\parallel} - 1\right) k_\parallel^2 v_{t_\parallel}^2 - \Omega_0^2}}{2} \tag{17}$$

### B. Semi-relativistic bi-Maxwellian Distribution

Semi-relativistic particle velocity distribution is believed to occur in some environments e.g., in the solar corona and magnetosheath where the perpendicular temperature dominates over the parallel. This distribution function is given by[42]

$$f_0 = \frac{1}{\sqrt{2\pi m T_\parallel}} \frac{\left(\frac{mc^2}{T_\perp}\right)}{(2\pi m T_\perp)} \left(1 + \frac{mc^2}{T_\perp}\right)^{-1} \exp\left[-\frac{mc^2}{T_\perp}\left(\sqrt{1 + \frac{p_\perp^2}{m^2 c^2}} - 1\right)\right] \exp\left[-\frac{p_\parallel^2}{2mT_\parallel}\right] \tag{18}$$

where $p_\perp = \gamma_\perp m v_\perp$ and $\gamma_\perp = \sqrt{1 + \frac{p_\perp^2}{m^2 c^2}}$.

After substituting $f_{0\perp}$ from Eq.[18] in Eq.[5] and assuming $|\omega| \ll ck_\parallel$, the dispersion relation for $\xi \ll 1$ becomes

$$\begin{aligned}
0 &= \frac{c^2 k_\parallel^2}{\omega_{pe}^2} + \left(\frac{c^2}{2mT_\perp^3}\right)\left(1 + \frac{mc^2}{T_\perp}\right)^{-1} \int_0^\infty dp_\perp \frac{p_\perp^3}{\gamma_\perp^2} \exp\left[-\frac{mc^2}{T_\perp}\left(\sqrt{1 + \frac{p_\perp^2}{m^2 c^2}} - 1\right)\right] \\
&\quad \times \left[\left(\frac{T_\perp}{T_\parallel} - 1\right) + i\sqrt{\pi}\frac{\omega}{k_\parallel v_{t_\parallel}}\left\{\frac{T_\perp}{T_\parallel} - \frac{\Omega_0/\gamma_\perp}{\omega}\left(\frac{T_\perp}{T_\parallel} - 1\right)\right\}\right]
\end{aligned} \tag{19}$$

To perform $p_\perp$-integration, we shall require the following integrals

$$
\begin{aligned}
(I_1, I_2, I_3) &= \int_0^\infty dp_\perp \frac{p_\perp^3}{\gamma_\perp^2} \exp\left[-\frac{mc^2}{T_\perp}\sqrt{1+\frac{p_\perp^2}{m^2c^2}}\right]\left(1, \frac{1}{\gamma_\perp}\right) \\
&= m^4 c^4 \int_1^\infty dx \frac{(x^2-1)}{x} \exp\left[-\alpha_\perp x\right]\left(1, \frac{1}{x}\right) \\
&= m^4 c^4 \{E_{-1}(\alpha_\perp) - E_1(\alpha_\perp),\ E_0(\alpha_\perp) - E_2(\alpha_\perp)\}
\end{aligned}
\tag{20}
$$

where $E_n(\alpha)$ is the standard exponential integral defined as

$$E_n(\alpha) = \int_1^\infty dx \frac{\exp[-\alpha x]}{x^n}$$

Resultantly, we obtain the real and imaginary parts of $\omega$ as

$$\operatorname{Re}\omega = \frac{\Omega_0}{\frac{T_\perp}{T_\parallel}}\left\{\left(\frac{T_\perp}{T_\parallel}-1\right)\right\}\frac{\chi_1}{\chi_2} \tag{21}$$

and

$$\operatorname{Im}\omega = \frac{k_\parallel v_{t_\parallel}}{\sqrt{\pi}}\left(\frac{T_\parallel}{T_\perp}\right)\left[\left(\frac{T_\perp}{T_\parallel}-1\right)-\frac{c^2 k_\parallel^2}{\omega_p^2}\frac{1}{\chi_2}\right] \tag{22}$$

where

$$\chi_1 = \frac{\left(\frac{mc^2}{T_\perp}\right)^3 \exp[\frac{mc^2}{T_\perp}]}{2\left(1+\frac{mc^2}{T_\perp}\right)}\{E_0(\alpha_\perp) - E_2(\alpha_\perp)\} = \frac{\alpha_\perp^2(1-\alpha_\perp)}{2(1+\alpha_\perp)}\left[1+\frac{e^{\alpha_\perp}\alpha_\perp^2 \Gamma(0,\alpha_\perp)}{1-\alpha_\perp}\right]$$

$$\chi_2 = \frac{\left(\frac{mc^2}{T_\perp}\right)^3 \exp[\frac{mc^2}{T_\perp}]}{2\left(1+\frac{mc^2}{T_\perp}\right)}\{E_{-1}(\alpha_\perp) - E_1(\alpha_\perp)\} = \frac{\alpha_\perp}{2}\left[1-\frac{e^{\alpha_\perp}\alpha_\perp^2 \Gamma(0,\alpha_\perp)}{1+\alpha_\perp}\right]$$

Here $\alpha_\perp = \frac{mc^2}{T_\perp}$ and $\Gamma(0,\alpha_\perp)$ is the standard incomplete gamma function.

The instability condition now becomes

$$k_\parallel^2 < \frac{\omega_p^2}{c^2}\left(\frac{T_\perp}{T_\parallel}-1\right)\chi_2 \tag{23}$$

Eqs.[21] & [22] describe the Weibel mode in the presence of the external magnetic field for a semi-relativistic electron plasma. $\chi s$ symbolize the relativistic effects, each $\chi$ approaches to unity in the non-relativistic limit and to zero in the highly relativistic limit. As is evident from Eqs. $[21-23]$, the relativistic effect expressed through $\chi s$ reduces the oscillations, enhances the damping term and thus suppresses the instability and also lowers the instability threshold in wavenumber.

Neglecting the ambient magnetic field yields the standard Weibel instability for relativistic perpendicular momentum distribution. Here the instability is suppressed due to the relativistic effect expressed through $\chi_2$. In an earlier work[42] on the subject, the authors missed Lorentz factor $\gamma$ in the semi-relativistic dispersion relation effecting the Im $\omega$ expression. Eq.[22] above correctly records the result.

We now derive the dispersion relation for $\xi \gg 1$. Substituting $f_{0\perp}$ from Eq.[18] in Eq.[6] and taking $|\omega| \ll ck_\parallel$, we obtain

$$0 = c^2 k_\parallel^2 - \frac{\omega_{pe}^2 \left(\frac{mc^2}{T_\perp}\right)}{2m^2 T_\perp^2 \left(1+\frac{mc^2}{T_\perp}\right)} \int_0^\infty \frac{dp_\perp \, p_\perp^3}{\gamma^2} \exp[-\frac{mc^2}{T_\perp}\left(\sqrt{1+\frac{p_\perp^2}{m^2 c^2}}-1\right)]$$
$$\times \left(\frac{\omega}{\omega-\Omega} + \left(\frac{T_\perp}{T_\parallel}-1\right)\frac{k_\parallel^2 v_{t_\parallel}^2}{2(\omega-\Omega)^2}\right) \qquad (24)$$

To facilitate $p_\perp$-integration, we assume $|\omega| \gg \Omega_0/\gamma_\perp$ so that the Eq.[24] becomes

$$0 = -\omega^2 + c^2 k_\parallel^2 + 2\pi \sum \frac{\omega_p^2}{m^2 v_{t\perp}^2} \int_0^\infty dp_\perp \frac{p_\perp^3 f_{o\perp}}{\gamma_\perp^2} \left\{\left(1+\frac{\Omega}{\omega \gamma_\perp}\right) + \left(\frac{T_\perp}{T_\parallel}-1\right)\frac{k_\parallel^2 v_{t_\parallel}^2}{2\omega^2}\left(1+\frac{2\Omega}{\omega \gamma_\perp}\right)\right\} \qquad (25)$$

or

$$0 = c^2 k_\parallel^2 + \sum \omega_p^2 \frac{\left(\frac{mc^2}{T_\perp}\right) exp[\frac{mc^2}{T_\perp}]}{2m^2 T_\perp^2 \left(1+\frac{mc^2}{T_\perp}\right)} \int_0^\infty dp_\perp \frac{p_\perp^3}{\gamma_\perp^2} exp\left[-\frac{mc^2}{T_\perp}\sqrt{1+\frac{p_\perp^2}{m^2 c^2}}\right]$$
$$\times \left\{\left(1+\frac{\Omega_0}{\gamma_\perp \omega}\right) + \left(\frac{T_\perp}{T_\parallel}-1\right)\frac{k_\parallel^2 v_{t_\parallel}^2}{2\omega^2}\left(1+\frac{2\Omega_0}{\gamma_\perp \omega}\right)\right\}$$

Using the results of integrals $I_1$ & $I_2$ from Eq.[20], we obtain

$$0 = c^2 k_\parallel^2 + \omega_p^2 \left\{\left(\chi_2 + \frac{\Omega_0}{\omega}\chi_1\right) + \left(\frac{T_\perp}{T_\parallel}-1\right)\frac{k_\parallel^2 v_{t_\parallel}^2}{2\omega^2}\left(\chi_2 + \frac{2\Omega_0}{\omega}\chi_1\right)\right\} \qquad (26)$$

where $\chi s$ are the same as in Eq.[22]. We may re-write Eq.[26] as

$$0 = \left(c^2 k_\parallel^2 + \omega_p^2 \chi_2\right)\omega^2 + \omega \Omega_0 \omega_p^2 \chi_1 + \frac{1}{2}\left(\frac{T_\perp}{T_\parallel}-1\right) k_\parallel^2 v_{t_\parallel}^2 \omega_p^2 \chi_2$$

The real frequency turns out to be unphysical, while the Im $\omega$ is given by

$$\text{Im}\,\omega = \omega_p \frac{\sqrt{2\left(\frac{T_\perp}{T_\parallel}-1\right) k_\parallel^2 v_{t_\parallel}^2 \chi_2 \left(c^2 k_\parallel^2 + \omega_p^2 \chi_2\right) - \Omega_0^2 \omega_p^2 \chi_1^2}}{2\left(c^2 k_\parallel^2 + \omega_p^2 \chi_2\right)} \qquad (27)$$

where the instability condition is

$$k_\parallel^2 > \chi_1 \frac{\omega_p^2}{2 c^2}\left[\left\{1 + \frac{2}{\left(\frac{T_\perp}{T_\parallel}-1\right)}\frac{c^2}{v_{t_\parallel}^2}\frac{\Omega_0^2}{\omega_p^2}\frac{\chi_2^2}{\chi_1^3}\right\}^{\frac{1}{2}} - 1\right] \qquad (28)$$

From Eq.[27], we note that as $\chi_2$ approaches to zero in the highly relativistic case, the instability is suppressed. The magnetic field effect is also reduced due to $\chi_1$. It may be noted that the relativistic effect also lowers the instability threshold.

In the limit of large anisotropy, the growth rate from Eq.[27] reduces to

$$\operatorname{Im}\omega = \frac{\sqrt{2\,k_\parallel^2 v_{t_\perp}^2 \omega_p^2 \chi_2 \left(c^2 k_\parallel^2 + \omega_p^2 \chi_2\right) - \Omega_0^2 \omega_p^4 \,\chi_1^2}}{2\left(c^2 k_\parallel^2 + \omega_p^2\,\chi_2\right)} \tag{29}$$

which is the modified form of Eq.[14] for the semi-relativistic velocity distribution. If we neglect the ambient magnetic field, we obtain the growth rate in semi-relativistic case as

$$\operatorname{Im}\omega = \frac{k_\parallel v_{t_\perp}}{\sqrt{2}} \frac{\omega_p \chi_2}{\sqrt{\left(c^2 k_\parallel^2 + \omega_p^2 \chi_2\right)}} \tag{30}$$

As the presence of $\chi_2$ shows, this result is the modified expression of Weibel's result[1], due to the semi-relativistic effect.

The non-relativistic results of Eq. $[16-17]$ for shorter and longer wavelengths are modified in the semi-relativistic case as

$$\operatorname{Im}\omega = \frac{\omega_p}{2}\sqrt{2\chi_2 \frac{v_{t_\parallel}^2}{c^2}\left(\frac{T_\perp}{T_\parallel}-1\right) - \frac{\Omega_0^2 \omega_p^2}{\left(c^4 k_\parallel^4\right)}\chi_1^2} \tag{31}$$

and

$$\operatorname{Im}\omega = \frac{\sqrt{2\left(\frac{T_\perp}{T_\parallel}-1\right) k_\parallel^2 v_{t_\parallel}^2 - \Omega_0^2 \left(\frac{\chi_1^2}{\chi_2^2}\right)}}{2} \tag{32}$$

for shorter wavelengths and longer wavelengths, respectively.

## III. Summary of Results and Discussion

Using the kinetic theory, we have derived a general electromagnetic dispersion relation for a magnetized anisotropic thermal plasma, with the propagation direction parallel to the background magnetic field. We have determined the analytical dispersion relations for both the non-relativistic and the semi-relativistic distributions under the limiting cases $\xi_\pm \leq 1$ and $\xi_\pm \gg 1$. In the non-relativistic case, we find that, due to the presence of ambient magnetic field the dispersion relation is no longer pure imaginary, as it also develops real oscillations for both the limiting cases $\xi_\pm \leq 1$ and $\xi_\pm \gg 1$.

In the semi-relativistic Maxwellian momentum distribution, while the parallel part is non-relativistic, the perpendicular part is relativistic which on implementation generates $\alpha = mc^2/T_\perp$ dependent terms including $\chi s$. These terms symbolize the relativistic effect. As $\alpha$ increases to larger values, each $\chi$ increases and eventually approaches unity i.e., the non-relativistic limit. On the other hand as $\alpha$ decreases towards smaller values, $\chi s$ also decrease and finally reach the ultra-relativistic limit[1,48].

In the semi-relativistic case, we observe that the growth rates and the real frequencies are suppressed due to the relativistic effect in both the limiting cases. The real frequency however becomes unphysical in the limit $\xi_\pm \gg 1$. The contribution of the external magnetic field is rendered insignificant in the limiting case $\xi_\pm \leq 1$. For $\xi_\pm \gg 1$, the magnetic field effect reduces the growth rate and diminishes as we move from the weak relativistic regime to the highly relativistic regime. Additionally, we observe from Eq.[23] and Eq.[28] that the instability threshold in wave number decreases as we move from non-relativistic to realtivistic regime due to the presence of $\chi s$ .

A graphical representation is also given for both the non-relativistic and the semi-relativistic cases. The imaginary part is plotted exhibiting the variation of $\text{Im}\,\omega\,/\,\omega_{pe}$ against $ck\,/\,\omega_{pe}$. For our graphical representation, we have used the plasma parameters[49]: $(i)\,\xi_{\pm}\,\gg\,1\,:\,\Omega_{0e}/\omega_{pe}\,=\,0.1\,,\,v_{t_{\parallel}}\,=\,0.01\,c$ and $T_{\perp}/T_{\parallel}\,=\,21$ ; $(ii)\,\xi_{\pm}\,\leq\,1\,:\,\Omega_{0e}/\omega_{pe}\,=\,0.01\,,\,v_{t_{\parallel}}\,=\,0.2\,c$ and $T_{\perp}/T_{\parallel}\,=\,3$. Figs. 1 & 2 display the imaginary part in the semi-relativistic case for different values of $\alpha$ for the limiting cases $\xi_{\pm}\,\leq\,1$ and $\xi_{\pm}\,\gg\,1$, respectively. Both the figures demonstrate that the growth rate is reduced due to the relativistic effect and that the threshold point is shifted towards lower value in $k_{\parallel}$. Our results in this paper can be useful for investigating Weibel instability with or without ambient magnetic field in a semi-relativistic environment which may occur in Nature or be created in the laboratory.

## IV. Research Bibliography

# List of Caption

Fig. 1. The $\frac{ck}{\omega_{pe}}$ vs $\frac{\text{Im}\,\omega}{\omega_{pe}}$ diagram for Eq.[22] .

Fig. 2. The $\frac{ck}{\omega_{pe}}$ vs $\frac{\text{Im}\,\omega}{\omega_{pe}}$ diagram for Eq.[27] .

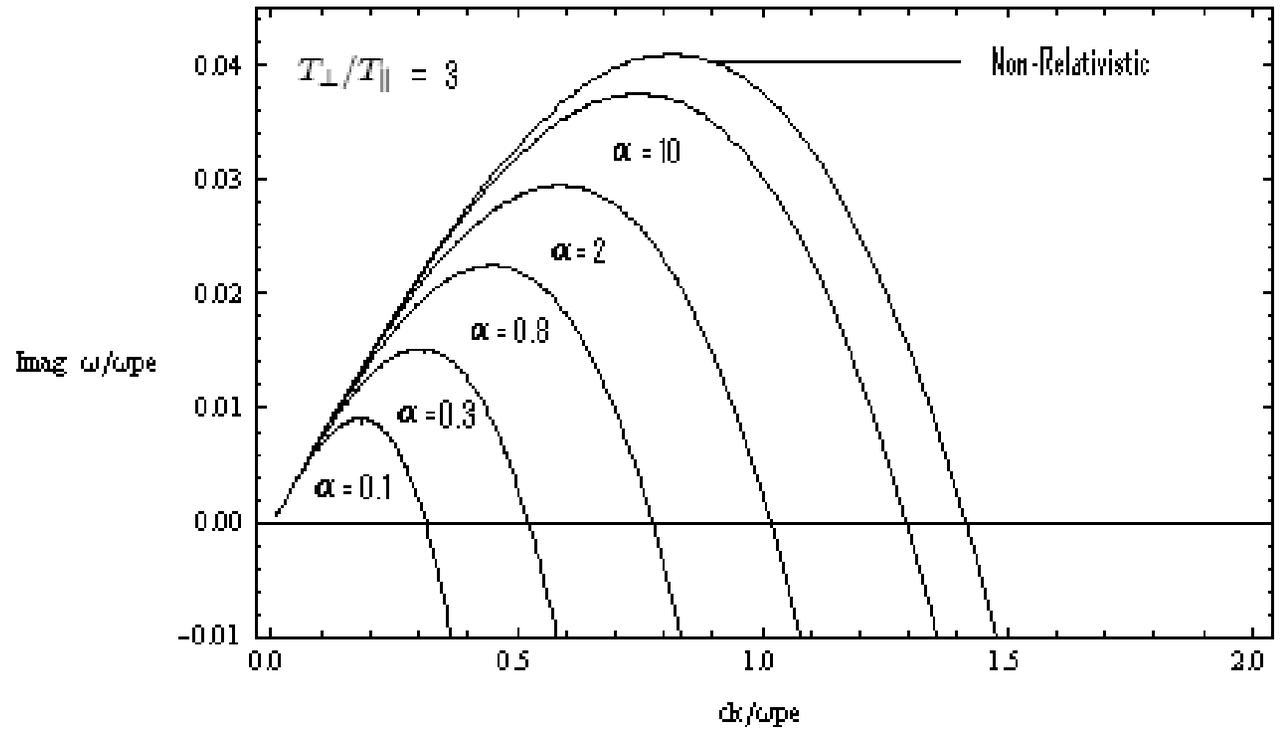

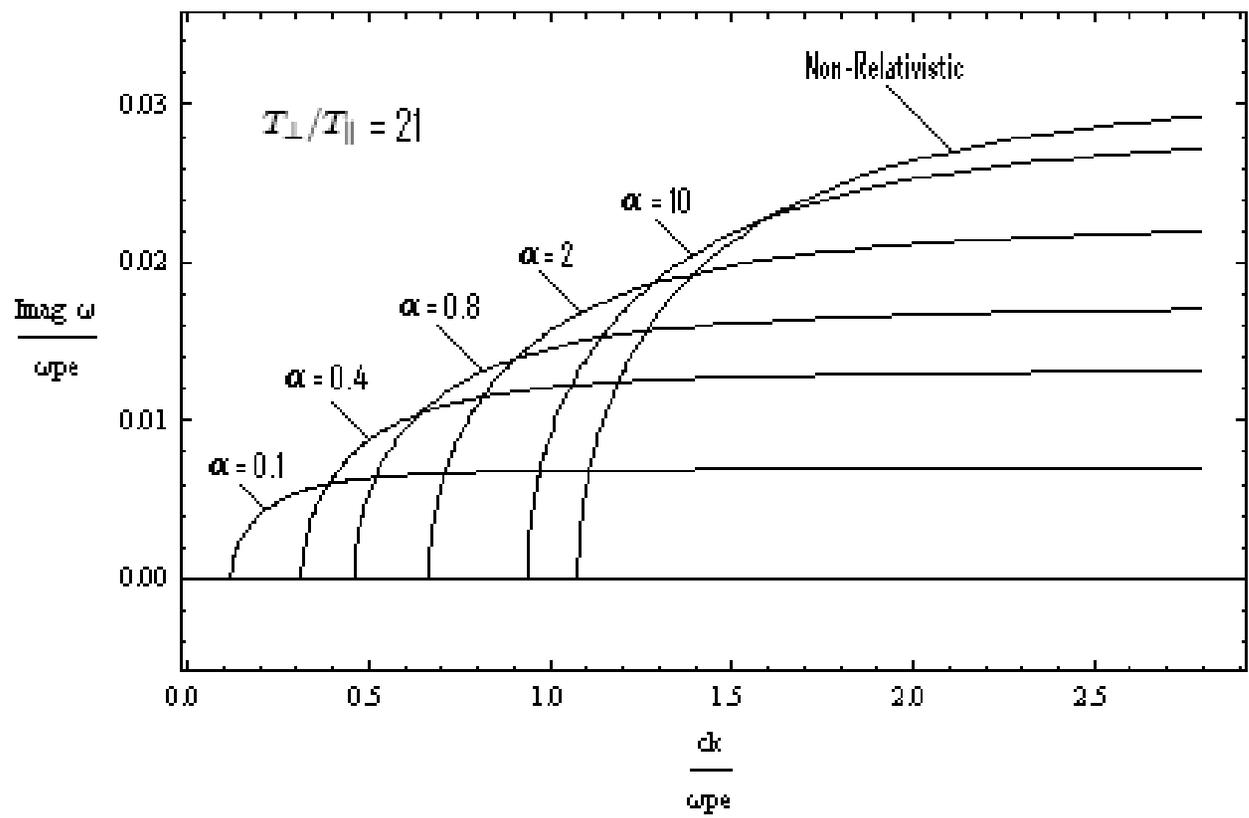

**Fig. 2**